\documentclass[a4paper,fleqn,usenatbib]{mnras}
\usepackage[T1]{fontenc}
\usepackage{ae,aecompl}
\usepackage{amsmath}
\usepackage{amsfonts}
\usepackage{amssymb}
\usepackage{graphicx}
\usepackage{color,xcolor}
\usepackage{tabularx}
\newcolumntype{L}{>{\raggedright\arraybackslash}X}
\def\mnras{MNRAS}
\def\aap{A$\&$A}
\def\nat{Nature}
\def\apj{ApJ}
\def\apjl{ApJ Letters}
\def\apjs{ApJS}
\def\pasp{PASP}
\def\aj{AJ}

\newcommand{\msun}{\, {\rm M}_{\odot}}
\newcommand{\msunyr}{\, {\rm M}_{\odot}\,{\rm yr^{-1}}}

\newcommand{\au}{\, {\rm au}}

\title[External photoevaporation microphysics]{How the microphysical properties of external photoevaporation influence the global evolution of protoplanetary discs}
\author[G. A. L. Coleman et al]{Gavin A. L. Coleman\thanks{Email: gavin.coleman@qmul.ac.uk}, Thomas J. Haworth and Lin Qiao\\
Astronomy Unit, Department of Physics and Astronomy, Queen Mary University of London, Mile End Road, London, E1 4NS, UK}
\date{Accepted 2025 April 2. Received 2025 March 19; in original form 2025 February 10}
\pubyear{2025}
\begin{document}
\label{firstpage}
\pagerange{\pageref{firstpage}--\pageref{lastpage}}
\maketitle
\begin{abstract}
External photoevaporation is one of the dominant mechanisms for mass loss from protoplanetary discs. However this mass loss is theoretically expected to depend upon the microphysical properties of protoplanetary discs, which are currently poorly constrained in observations. In this work we explore the impact of microphysics on the bulk evolution of discs. The polycyclic aromatic hydrocarbon (PAH) abundance, and the extent to which grain growth has occurred in the disc have profound effects on the strength of mass loss rates due to external photoevaporation, which in turn can have a significant impact on the disc evolution, impacting disc radii and accretion rates over time. The strongest sensitivity is to whether grain growth has occurred in the disc, which reduces the amount of dust entrained in the wind to shield the disc, thus increasing the rate at which gas is lost. Additionally, larger PAH abundances result in stronger heating and higher mass loss rates, but to a lesser extent than grain growth. We find that plausible variations in the PAH abundance and disc dust evolution can leave observable differences in disc populations. This work highlights the importance of obtaining observational constraints of the microphysical properties of protoplanetary discs. Future observations from JWST should soon be able to provide these constraints.

\end{abstract}

\section{Introduction}
\label{sec:intro}

With planets thought to form in protoplanetary discs surrounding young pre-main sequence stars \citep{2018ApJ...869L..41A, 2018A&A...617A..44K, 2018ApJ...860L..13P, 2019Natur.574..378T}, understanding the evolution of such discs is of utmost importance. Thanks primarily to {\it ALMA} and {\it VLT/SPHERE}, a large range of individual features, both axisymmetric (e.g. rings) and non-axisymmetric (e.g. spirals), have been observed \citep[for a recent review see][]{Andrews20}. Simultaneously, numerous statistics on fundamental disc properties such as disc masses/radii \citep{Miotello23} have also been generated.
Additionally, numerous works have explored the diversity in disc lifetimes across different clusters of varying ages \citep{Haisch01,Ribas15,Pfalzner22}, and more recently across populations of similar mass stars \citep{Pfalzner24}.
With all of these features and properties varying in time and space over the lifetime of protoplanetary discs, this too adds incentives to understand how protoplanetary discs evolve.

The evolution of protoplanetary discs occurs through a range of mechanisms. For example discs evolve through accreting on to the central star \citep[e.g.][]{1981ARA&A..19..137P, 2016A&A...591L...3M, 2020A&A...639A..58M, 2021A&A...650A.196M}, internal photoevaporative \citep[e.g.][]{Alexander07, Owen12, 2017RSOS....470114E} or magnetically driven winds \citep{BalbusHawley1991, 2007prpl.conf..277P, Suzuki09, 2014prpl.conf..411T, 2016ApJ...818..152B, 2019ApJ...874...90W, Tabone22, Pascucci23} as well as winds driven by external irradiation from nearby massive stars \citep[e.g.][]{1994ApJ...436..194O, 2000ApJ...539..258R, 2004ApJ...611..360A, 2005AJ....130.1763S, 2016ApJ...826L..15K,  Haworth19, 2021MNRAS.501.3502H, 2022EPJP..137.1132W, PastPresentFuture2025}. Signatures of all of these processes have been observed in protoplanetary discs \citep[e.g.][]{1994ApJ...436..194O, 2016A&A...591L...3M, 2017RSOS....470114E, 2018A&A...609A..87N, 2021ApJS..257...16B}.
Various combinations of these processes are also routinely used in disc evolution studies \citep[e.g.][]{Coleman20,Coleman22}, whilst also being important for the types and frequencies of planets that are able to form in protoplanetary discs \citep{Winter22,Qiao23,Coleman24,Coleman24FFP}.

The main test for disc evolution models is how well are they able to match or provide predictions for observations. Typically, this involves comparisons with the statistical properties of large numbers of protoplanetary discs. For example \citet{Alexander23} compared analytic solutions for wind driven discs and viscous discs, including internal photoevaporation, concentrating on mass accretion rates, and finding that sufficiently large observational datasets could differentiate between the models. Comparing the disc mass and mass accretion rates together, it was previously expected that there would be a linear correlation for viscous discs \citep{Jones12,Lodato17}, however more recent work including internal photoevaporation predicts that a ``knee'' feature would appear when the mass accretion rate fell below the photoevaporative mass loss rate \citep{Somigliana20}. The other main observable that is compared between observations and theoretical studies is the disc radius, and by proxy the disc mass. Previous work including external photoevaporation in viscous discs found that the outer disc edge is balanced when the viscous expansion rate equals the mass loss rate \citep{Clarke07}. Other works comparing observed disc radii to theoretical models propose that observed correlations between disc radii and dust fluxes is a signature of grain growth and points to low levels of viscosity \citep{Rosotti19b}. The impact of substructure on the evolution of externally photoevaporating discs has also been studied by \cite{2024A&A...681A..84G} who find that dust pressure traps can extend the disc radius and lifetime. 

Within our own models for disc evolution, \citet{Coleman22} included both internal and external photoevaporation within viscously evolving discs and found multiple evolution pathways that protoplanetary discs follow. When comparing to observations of disc lifetimes in stellar clusters, they found significant degeneracy between parameters when trying to match to observations. More recently, \citet{Coleman24MHD} compared the evolution of viscous and wind driven discs using more uptodate prescriptions for external photoevaporation. They found that photoevaporation is detrimental to determining the differences between viscous and MHD wind driven discs since either internal or external photoevaporation dominate the mass loss rates of the discs. The latter work utilises the new updated {\sc fried} grid of externally driven photoevaporative mass loss rates \citep{Haworth23}. This grid provides external photoevaporative mass loss rates tabulated over a wide range of external UV field strengths, stellar masses, disc masses and disc radii. It includes many improvements on the original grid \citep{Haworth18}, but most notably varies the microphysics of the photodissociation hydrodynamics calculations to i) permit variations in the polycyclic aromatic hydrocarbon (PAH) abundance, which is the dominant heating mechanism and ii) to switch between ISM-like and evolved dust in the outer disc, which massively affects the grain entrainment and extinction in the wind \citep{2016MNRAS.457.3593F}. Combinations of those microphysics options can change the mass loss rate by an order of magnitude or more in some cases.

In this work, we include the new {\sc fried} grid of externally driven mass loss rates \citep{Haworth23}, within our viscously evolving disc models that already include prescriptions for internal photoevaporation \citep{Picogna19,Ercolano21}. Our aim here is to determine the differences in the  microphysics of external photoevaporation affect the evolution of protoplanetary disc models.  We find that the microphysics does indeed matter and does affect the large scale evolution of the disc. This introduces further degeneracy when comparing disc evolutionary models with observations of protoplanetary discs. Ultimately the results highlight the importance of having tighter constraints on the PAH abundance and dust properties within externally irradiated discs and their winds, of which facilities such as {\it JWST} are ideally placed to produce. 

This paper is organised as follows.
Section \ref{sec:physical_model} outlines the disc evolution and photoevaporation models as well as the simulation parameters.
In sect. \ref{sec:results} we present the results of our disc evolution models, highlighting the effects of different mass loss mechanisms.
Finally, we draw our conclusions in sect. \ref{sec:conclusions}.

\section{Physical Model and Parameters}
\label{sec:physical_model}

Protoplanetary discs lose mass by accretion onto the central star and through photoevaporative winds launched from the disc surface layers.
To account for these processes we adopt a 1D viscous disc model similar to that used in previous works \citep{ColemanNelson14,ColemanNelson16,Coleman21,Coleman22} where the equilibrium temperature is calculated by balancing irradiation heating from the central star, background heating from the residual molecular cloud, viscous heating and blackbody cooling.
The surface density, $\Sigma$, is evolved by solving the standard diffusion equation
\begin{equation}
    \dot{\Sigma}(r)=\dfrac{1}{r}\dfrac{d}{dr}\left[3r^{1/2}\dfrac{d}{dr}\left(\nu\Sigma r^{1/2}\right)\right]-\dot{\Sigma}_{\rm PE}(r)
\end{equation}
where $\nu=\alpha H^2\Omega$ is the disc viscosity with viscous parameter $\alpha$ \citep{Shak}, $H$ being the disc scale height, $\Omega$ the Keplerian frequency, and $\dot{\Sigma}_{\rm PE}(r)$ is the rate of change in surface density due to photoevaporative winds.
Following \citet{Coleman22} we include EUV and X-ray internal photoevaporative winds from the central star (detailed in section \ref{sec:internalPhoto}) as well as winds launched from the outer disc by far ultraviolet (FUV) radiation emanating from nearby massive stars (e.g. O-type stars, see section \ref{sec:externalPhoto}).
We assume that the photoevaporative mass loss rate at any radius in the disc is the maximum of the internally and externally driven rates 
\begin{equation}
    \dot{\Sigma}_{\rm PE}(r) ={\rm max}\left(\dot{\Sigma}_{\rm I,X}(r),\dot{\Sigma}_{\rm E,FUV}(r)\right)
\end{equation}
where the subscripts I and E refer to contributions from internal and external photoevaporation.

\subsection{Internal Photoevaporation}
\label{sec:internalPhoto}
The absorption of high energy radiation from the host star by the disc can heat the gas above the local escape velocity, and hence drive internal photoevaporative winds. EUV irradiation creates a layer of ionised hydrogen with temperature $\sim$10$^4$~K \citep{Clarke2001}, however X-rays penetrate deeper into the disc and are still capable of heating up to around $\sim$10$^4$~K \citep{Owen10} so for low mass stars are expected to generally dominate over the EUV for setting the mass loss rate. FUV radiation penetrates deeper still, creating a neutral layer of dissociated hydrogen with temperature of roughly 1000K \citep{Matsuyama03}. The overall interplay between the EUV, FUV and X-rays is a matter of ongoing debate.  \cite{Owen12} find that including the FUV heating simply causes the flow beneath the sonic surface to adjust, but otherwise retains the same mass loss rate. However others suggest a more dominant role of the FUV \citep[e.g.][]{2009ApJ...705.1237G,2015ApJ...804...29G}. Recent models including all three fields suggest a more complicated interplay \citep[e.g.][]{2017ApJ...847...11W, 2018ApJ...865...75N,Ercolano21}.  The outcome also depends sensitively on how the irradiated spectrum is treated \citep{2022MNRAS.514..535S}. 

The radiation hydrodynamic models of \cite{Owen12} used pre-computed X-ray driven temperatures as a function of the ionisation parameter ($\xi = L_X / n /r^2$) wherever the column to the central star is less than $10^{22}$cm$^{-2}$ (and hence optically thin). This approach has since been updated with a series of column-dependent temperature prescriptions \citep{Picogna19,Ercolano21,Picogna21}.  

We follow \citet{Picogna21} who further build on the work of \cite{Picogna19} and \cite{Ercolano21} and find that the mass loss profile from internal X-ray irradiation is approximated by
\begin{equation}
\label{eq:sig_dot_xray}
\begin{split}
\dot{\Sigma}_{\rm I,X}(r)=&\ln{(10)}\left(\dfrac{6a\ln(r)^5}{r\ln(10)^6}+\dfrac{5b\ln(r)^4}{r\ln(10)^5}+\dfrac{4c\ln(r)^3}{r\ln(10)^4}\right.\\
&\left.+\dfrac{3d\ln(r)^2}{r\ln(10)^3}+\dfrac{2e\ln(r)}{r\ln(10)^2}+\dfrac{f}{r\ln(10)}\right)\\
&\times\dfrac{\dot{M}_{\rm X}(r)}{2\pi r} \dfrac{\msun}{\au^2 {\rm yr}}
\end{split}
\end{equation}
where
\begin{equation}
\label{eq:m_dot_r_xray}
    \dfrac{\dot{M}_{\rm X}(r)}{\dot{M}_{\rm X}(L_{X})} = 10^{a\log r^6+b\log r^5+c\log r^4+d\log r^3+e\log r^2+f\log r+g}
\end{equation}
where $a=-0.6344$, $b=6.3587$, $c=-26.1445$, $d=56.4477$, $e=-67.7403$, $f=43.9212$, and $g=-13.2316$.
We follow \citet{Komaki23} and apply a simple approximation to the outer regions of the disc where the internal photoevaporation rates drop to zero.
In models of internal photoevaporation, it is assumed that the wind from the inner regions of the disc blocks radiation from heating the outer regions and being able to drive winds from that location. This normally results in the sudden drop in the internal photoevaporation rate. However these models do not take into account the effects of when the disc and/or the wind become optically thin and therefore ineffective at blocking the radiation, and so in order to model the effects of the wind in those locations it is necessary to apply a simple approximation to the profiles presented by \citet{Picogna21}. An example of such a profile is shown in Fig. 9 of \citet{Komaki23} whilst we apply a similar profile here in the outer regions of the disc where the internal photoevaporative winds begin to drop to zero.
As the temperature of X-ray irradiated gas varies from $\sim 10^3$--$10^4$ K depending on the distance in the disc \citep[e.g.][]{Owen10}, we conservatively define the radius at which the internal photoevaporation scheme drops off as the gravitational radius for $10^3$ K gas.
We therefore apply the following approximation at radial distances greater than $r_{\rm rgx}$\footnote{Note that we use $r_{\rm rgx}$ as an arbitrary value, and a different radius in the disc would also be suitable as long as it is sufficiently distant from where internal photoevaporation mostly operates.}
\begin{equation}
    \dot{\Sigma}_{\rm I,X,ap} = \dot{\Sigma}_{\rm rgx}\left(\frac{r}{r_{\rm rgx}}\right)^{-1.578}
\end{equation}
where $\dot{\Sigma}_{\rm rgx}$ is equal to eq. \ref{eq:sig_dot_xray} calculated at $r=r_{\rm rgx}$, and
\begin{equation}
    r_{\rm rgx} = \dfrac{GM_*}{c_{\rm s}^2}
\end{equation}
where $c_{\rm s}$ is the sound speed for gas of temperature $T=10^3 K$, and $\mu=0.5$. For solar mass stars, the gravitational radius $r_{\rm rgx} \sim 85 \au$.
In the outer regions of the disc the loss in gas surface density due to internal photoevaporation then becomes
\begin{equation}
    \dot{\Sigma}_{\rm I}(r) = \max(\dot{\Sigma}_{\rm I,X}(r),\dot{\Sigma}_{\rm I,X,ap}).
\end{equation}

Following \cite{Ercolano21} the integrated mass-loss rate, dependant on the stellar X-ray luminosity, is given as
\begin{equation}
    \log_{10}\left[\dot{M}_{X}(L_X)\right] = A_{\rm L}\exp\left[\dfrac{(\ln(\log_{10}(L_X))-B_{\rm L})^2}{C_{\rm L}}\right]+D_{\rm L},
\end{equation}
in $\msunyr$, with $A_{\rm L} = -1.947\times10^{17}$, $B_{\rm L} = -1.572\times10^{-4}$, $C_{\rm L} = -0.2866$, and $D_{\rm L} = -6.694$.

\subsection{External Photoevaporation}
\label{sec:externalPhoto}

In addition to internal winds driven by irradiation from the host star, winds can also be driven from the outer regions of discs by irradiation from external sources. Massive stars dominate the production of UV photons in stellar clusters and hence dominate the external photoevaporation of discs. External photoevaporation has been shown to play an important role in setting the evolutionary pathway of protoplanetary discs \citep{Coleman22},their masses \citep{2014ApJ...784...82M, 2017AJ....153..240A}, radii \citep{2018ApJ...860...77E,Coleman24MHD} and lifetimes \citep{2016arXiv160501773G, 2019MNRAS.490.5678C, 2020MNRAS.492.1279S, 2020MNRAS.491..903W} even in weak UV environments \citep{2017MNRAS.468L.108H, 2023A&A...673L...2V}.
We do not include shielding of the protoplanetary discs, i.e. by the nascent molecular cloud, that has been shown to have an effect on the effectiveness of external photoevaporation \citep{Qiao22,Qiao23,Wilhelm23}, but instead will study constant weaker UV environments.
In our simulations, the mass loss rate due to external photoevaporation is calculated by interpolating over the recently updated \textsc{fried} grid \citep{Haworth23}, described further below and in Sect. \ref{sec:subgrids}.

We evaluate the \textsc{fried} mass loss rate at each radius from the outer edge of the disc down to the radius that contains 80$\%$ of the disc mass. We choose this value as 2D hydrodynamical models show that the vast majority of the mass loss from external photoevaporation, comes from the outer 20\% of the disc \citep{Haworth19}. The change in gas surface density is then calculated as
\begin{equation}
    \dot{\Sigma}_{\textrm{ext, FUV}}(r) = G_{\rm sm} \frac{\dot{M}_\textrm{{ext}}(R_{\textrm{\textrm{max}}})}{\pi(R^2_\textrm{{d}} - {R_{\textrm{\textrm{max}}}}^2)+A_{\rm sm}}, 
\end{equation}
where $A_{\rm sm}$ is a smoothing area equal to 
\begin{equation}
A_{\rm sm} = \dfrac{\pi(R_{\rm max}^{22}-(0.1 R_{\rm max})^{22})}{11R_{\rm max}^{20}}
\end{equation}
and $G_{\rm sm}$ is a smoothing function
\begin{equation}
    G_{\rm sm} = \dfrac{r^{20}}{R_{\rm max}^{20}}.
\end{equation}
In the above equations, $R_{\rm d}$ is the disc outer radius and $R_{\rm max}$ is the radial location where the mass loss rate is at a maximum.

\subsubsection{{\sc fried} subgrids}
\label{sec:subgrids}

One of the main features with the new {\sc fried} models is the development of different grids that take into account different ranges for the microphysics involved.
These are namely the abundance of PAHs and the question of whether substantial grain growth has occurred in the disc. Only small grains are entrained in an external photoevaporative wind \citep{2016MNRAS.457.3593F}, so if grain growth has occurred in the outer disc the small dust reservoir is depleted, less dust is entrained in the wind and the extinction to the disc outer edge is reduced. Grain growth therefore helps to enhance the external photoevaporative mass loss rate by reducing attenuation of the UV incident on the disc. The PAH abundance is important too because they are the dominant heating contributor in the wind launching region \citep{Haworth23} but their abundance is highly uncertain with only limited observational constraints.
Only one such system, HST 10, has a constraint on the PAH-to-gas ratio (1/50th that of the surrounding Orion bar \citep{Vicente13} and 1/90th that of the NGC 7023 cluster \citep{Berne12}). So whilst there is some idea that the PAH-to-gas abundance is probably depleted, we don't have a good idea of a fiducial value, or to what extent this is due to dust or PAH depletion.
For this reason, the {\sc fried} grid explored a range of PAH-to-dust ratios.

Given the above,  {\sc fried} had two options for the dust (ISM-like dust or grain growth) and three values of the PAH abundance. The PAH abundance can be defined relative to the gas, or relative to the dust. {\sc fried} controls it as a PAH-to-Dust ratio $f_d$ relative to a fiducial PAH-to-dust ratio in the ISM based on \cite{2003ApJ...587..278W}, and uses $f_d=0.1, 0.5, 1.0$. So if the dust option is ISM-like and $f_d=1$ the PAH abundance is exactly like \cite{2003ApJ...587..278W}, if the dust option is ISM-like and $f_d=0.1$ it is a factor 10 less abundant in PAHs. When grain growth is included the dust-to-gas mass ratio in the wind is further depleted by a factor 100 and so while the PAH-to-dust ratio is still defined by the three choices of $f_d$, the PAH-to-gas ratio would also be depleted by a factor 100. Throughout the rest of this paper we use a shorthand notation where the first term is $f_d$ and the second term is ``S'' or ``G'' depending on whether the dust is ISM like or has undergone grain growth. For example 0.1--S has a PAH-to-dust ratio of 0.1 of that used by \cite{2003ApJ...587..278W} and ISM-like dust. Typically the \textsc{fried} models with grain growth and higher PAH abundances have higher mass loss rates. These six subgrids can be found in Table \ref{tab:fried}.

\begin{table}
    \centering
    \begin{tabular}{cccc}
    \hline
        Subgrid & PAH--to--dust & PAH--to--gas & Grain growth\\
        & relative to ISM & relative to ISM & \\
        \hline
       1--G & 1 & 1/100 & on\\
       0.5--G & 0.5 & 1/200 & on\\
       0.1--G & 0.1 & 1/1000 & on\\
       1--S & 1 & 1 & off\\
       0.5--S & 0.5 & 1/2 & off\\
       0.1--S & 0.1 & 1/10 & off\\
        \hline
    \end{tabular}
    \caption{Different {\sc fried} subgrids}
    \label{tab:fried}
\end{table}
For the bulk of this work, we concentrate on two subgrids, $f_{\rm PAH,d}=1$ with grain growth (1--G), and $f_{\rm PAH,d}=0.1$ and ISM-like (0.1--S), found in \citet{Haworth23} to yield the strongest and weakest mass loss rates respectively.

\subsection{Simulation Parameters}

In this work, we only examine protoplanetary discs around Solar mass stars.
In addition to exploring the effects of the different subgrids of {\sc fried}, we also vary the viscous parameter $\alpha$ that accounts for the transport of material through a disc and on to the central star, the stellar X-ray luminosity that controls the internal photoevaporation rate, and the external FUV field strength that determines the external photoevaporation rate.

For the external photoevaporative mass loss rates, we vary the strength of the local environment, ranging from 10 $\rm G_0$ to $10^{5} \rm G_0$, appropriate for most star forming regions, e.g. Taurus or Orion.
X-ray luminosities are observed to vary by up to two orders of magnitude even for stars of the same mass, due to a combination of measurement uncertainty and genuine intrinsic differences in X-ray activity levels, which can be time varying \citep[see figure 1 of][ and the associated discussion]{Flaischlen21}.
We therefore account for this spread by considering X-ray luminosities with $L_X = 10^{29.5}$--$10^{31.5} {\rm erg s^{-1}}$ with the average being motivated by observations \citep{Flaischlen21}.

We initialise our disc following \citet{Lynden-BellPringle1974}
\begin{equation}
    \Sigma = \Sigma_0\left(\frac{r}{1\au}\right)^{-1}\exp{\left(-\frac{r}{r_{\rm C}}\right)}
\end{equation}
where $\Sigma_0$ is the normalisation constant set by the total disc mass, (for a given $r_{\rm C}$), and $r_{\rm C}$ is the scale radius, which sets the initial disc size, taken here to be equal to 50 $\au$.
For the initial mass of the disc we follow \citet{Haworth20} where from hydrodynamic simulations they find the maximum disc mass $M_{\rm d, max}$ that a gas disc of radius $r_{\rm ini}$ around a star of mass $M_*$ can be before becoming gravitationally unstable is equal to
\begin{equation}
    \label{eq:max_disc_mass}
    \dfrac{M_{\rm d, max}}{M_*} < 0.17 \left(\dfrac{r_{\rm ini}}{100\au}\right)^{1/2}\left(\dfrac{M_*}{\msun}\right)^{-1/2}.
\end{equation}
In this work, we assume that for eq. \ref{eq:max_disc_mass} the initial disc radius is equal to 100$\au$ and take the mass of the discs to be equal to $M_{\rm d, max}$, which corresponds to $M_{\rm d}=0.17\msun$.
Table \ref{tab:parameters} shows the simulation parameters that we used in this work.

\begin{table}
    \centering
    \begin{tabular}{c|c}
    \hline
    Parameter & 1 $\msun$ \\
    \hline
        $r_{\rm ini} (\au)$ & 100 \\
        $M_{\rm d,max} (\msun)$ & 0.17 \\
        $\alpha$ & $10^{-3.5}$ \\
        $L_X (\log_{10} (\rm erg~s^{-1}))$ & 29.5--31.5 \\
        UV Field ($\rm G_0)$ & $10^1$--$10^5$ \\
        $f_{\rm pah}$ & 0.1, 0.5, 1 \\
    \hline
    \end{tabular}
    \caption{Simulation Parameters for the models discussed in this paper.}
    \label{tab:parameters}
\end{table}

\section{Results}
\label{sec:results}

\begin{figure}
\centering
\includegraphics[scale=0.4]{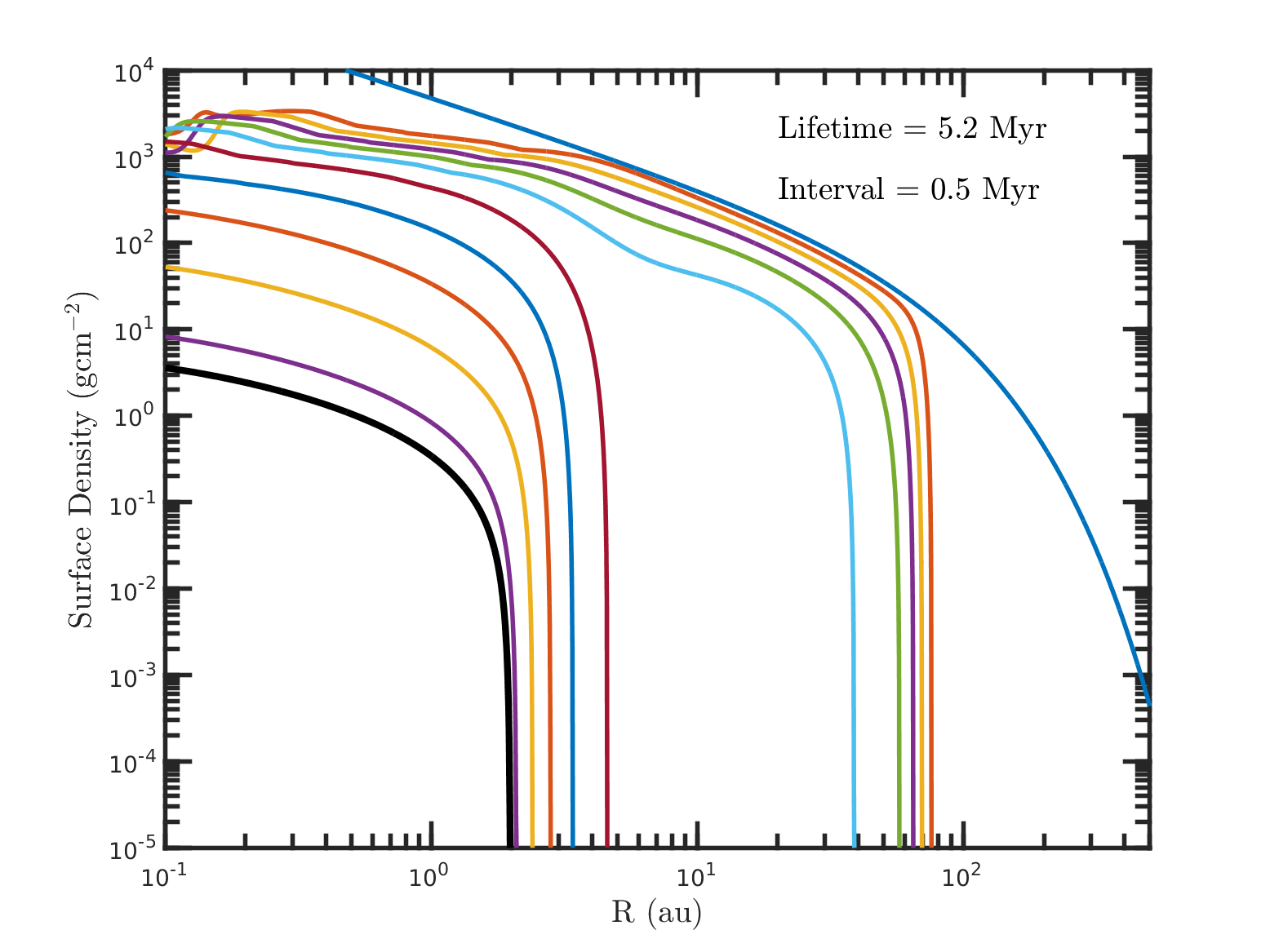}
\caption{Surface density profiles of an evolving protoplanetary disc with $\alpha=3\times 10^{-4}$, located in an $10^3 \rm G_0$ environment, and with a central X-ray luminosity $L_{\rm X}=3.2\times 10^{30}\rm erg s^{-1}$. The subgrid of {\sc fried} used is the $\rm f_{PAH}=1$ and that including grain growth. The uppermost blue line shows the initial disc profile, whilst the black line shows the last disc output before the disc had fully dispersed. The disc had a lifetime of 5.2 Myr and the surface density profiles are shown at intervals of 0.5 Myr.}
\label{fig:disc_1000G0_1G}
\end{figure}

We begin our results by showing the evolution of a protoplanetary disc using our fiducial model. Figure \ref{fig:disc_1000G0_1G} shows the temporal evolution of the gas surface density of a disc where the viscous parameter $\alpha = 3\times10^{-4}$, the central X-ray luminosity $L_{\rm X}= 10^{30.5} {\rm erg~s^{-1}}$, and the disc is placed in an external environment equal to $10^3 \rm G_0$. We use the subgrid in {\sc fried } that corresponds to $f_{\rm pah}=1$ and includes the effect of grain growth. The lifetime of the disc was equal to 5.2 Myr, and each profile in Fig. \ref{fig:disc_1000G0_1G} shows the gas surface density at 0.5 Myr intervals, with the blue profile in the top right of the figure being the initial profile, and the black line in the bottom left of the figure showing the final output.
The disc evolved from the top right to the bottom left.
After $\sim$ 0.5 Myr, the outer edge of the disc has truncated down to 75 $\au$, as shown by the red profile.
At that radius, viscous spreading matches the mass lost through photoevaporation and the disc radius stays roughly steady, truncating only slightly further over the next 1.5 Myr.
After $\sim 2.7$ Myr, the outer disc has dispersed sufficiently for internal photoevaporation to open a small gap in the disc at $\sim 15 \au$.
The disc exterior to this location, then dispersed within the next 0.3 Myr, leaving the disc now truncated to $\sim 5\au$. With the viscous time-scale being $\sim 2$Myr from $\sim$ few$\au$, the inner disc then slowly accreted on to the central star over the next 2 Myr.
This evolution was typical for the discs in our simulations, with the speed of the truncation being determined by the external photoevaporation rates, and the presence of a gap in the disc being determined by the strength of the central stars X-ray luminosity.
Essentially, the discs truncated down to a small size, where internal photoevaporation may have opened a gap in the discs, that then allowed the outer disc to be dispersed leaving an inner disc that viscously accreted onto the central star.
We will now discuss the effects that the choice of the {\sc fried} subgrid, the strength of the local environment, and the central stars X-ray luminosity have on the evolution of protoplanetary discs.

\subsection{Usage of different subgrids - PAH-to-dust ratios and grain growth}

\begin{figure}
\centering
\includegraphics[scale=0.4]{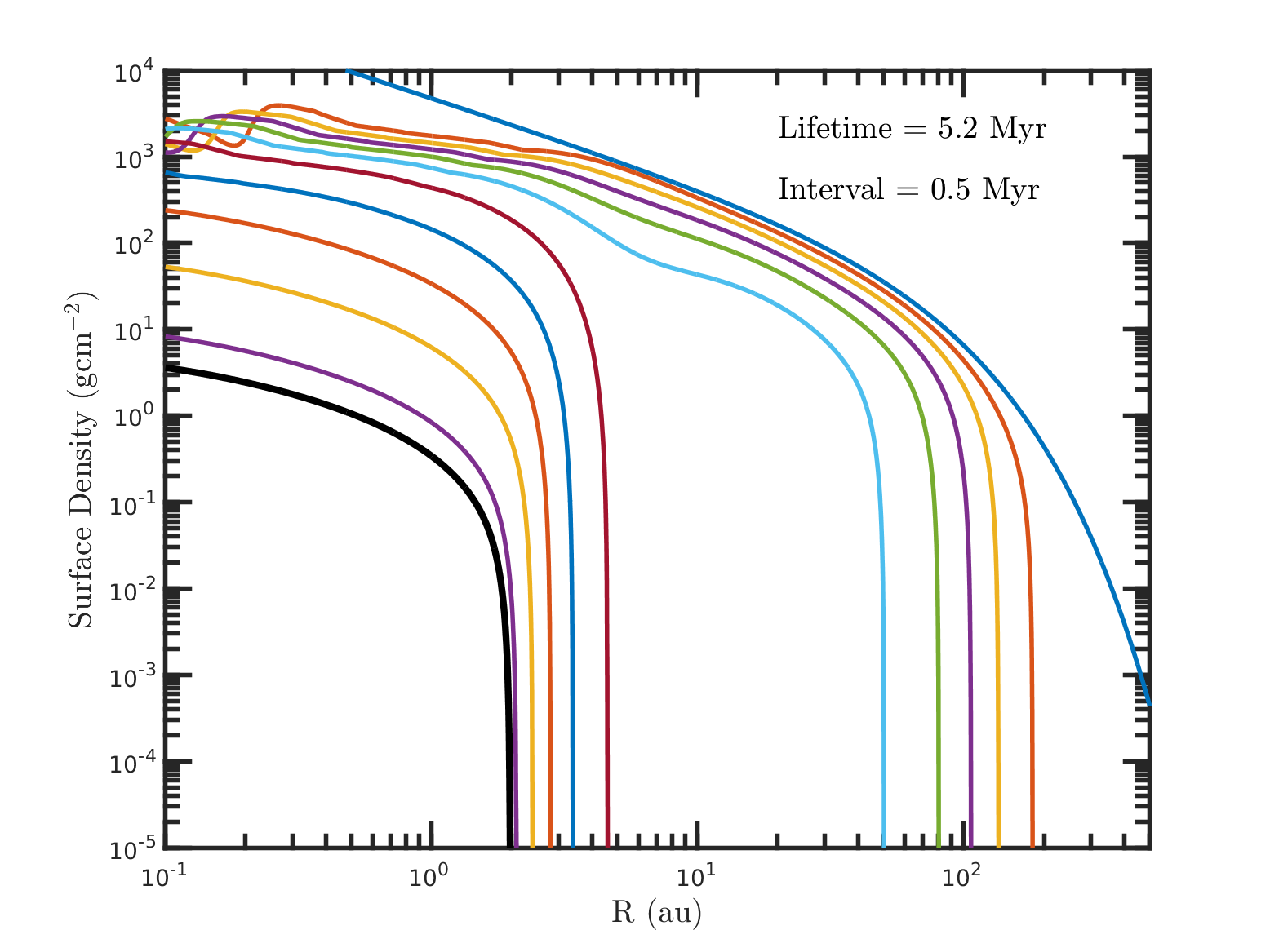}
\caption{Same as Fig. \ref{fig:disc_1000G0_1G} but for the {\sc fried} subgrid of $\rm f_{PAH}=0.1$ and of only ISM-like dust.}
\label{fig:disc_1000G0_0_1S}
\end{figure}

As stated in Sect. \ref{sec:subgrids}, one of the main features with the new {\sc fried} models is the development of different grids that take into account different ranges for the microphysics involved. We now explore the effects of changing subgrids by comparing the evolution of protoplanetary discs in different environments, whilst using the different subgrids that represent whether grain growth has occurred, and what the PAH abundance is.
In Fig. \ref{fig:disc_1000G0_0_1S} we show the evolution of the gas surface density of a disc in an intermediate UV environment, the same conditions as in Fig. \ref{fig:disc_1000G0_1G}. The difference here between Figs. \ref{fig:disc_1000G0_0_1S} and \ref{fig:disc_1000G0_1G} being that the {\sc fried} subgrid used in Fig. \ref{fig:disc_1000G0_0_1S} was the 0.1--S, a dusty, PAH-depleted wind, which yields the weakest mass loss rates \citep{Haworth23}.
When comparing Fig. \ref{fig:disc_1000G0_0_1S} to Fig. \ref{fig:disc_1000G0_1G}, even though the disc lifetimes were similar, $\sim$ 5.2 Myr, the evolution of the protoplanetary discs was notably different.
With the subgrid 1--G in Fig. \ref{fig:disc_1000G0_1G}, the disc was quickly truncated down to $\le75\au$ after less than 0.5 Myr.
However in the weaker subgrid 0.1--S, the disc only truncated to $\sim 200\au$ after 0.5 Myr and then slowly truncated down to $\sim 100\au$ over the next 1.5 Myr, and to $\sim50\au$ after 2.5 Myr.
Thus it is clearly evident that the reduction in the mass loss rate due to the different subgrid of {\sc fried} resulted in the disc remaining much more extended for at least half of the disc lifetime.
Once the disc truncated down to small radii ($\sim 50\au$), the evolution of the discs was similar, where internal photoevaporation was able to open a gap in the disc and subsequently disperse the outer disc, followed by the remaining lifetime of the disc being dependent on the strength of turbulence in the inner regions.

\begin{figure}
\centering
\includegraphics[scale=0.4]{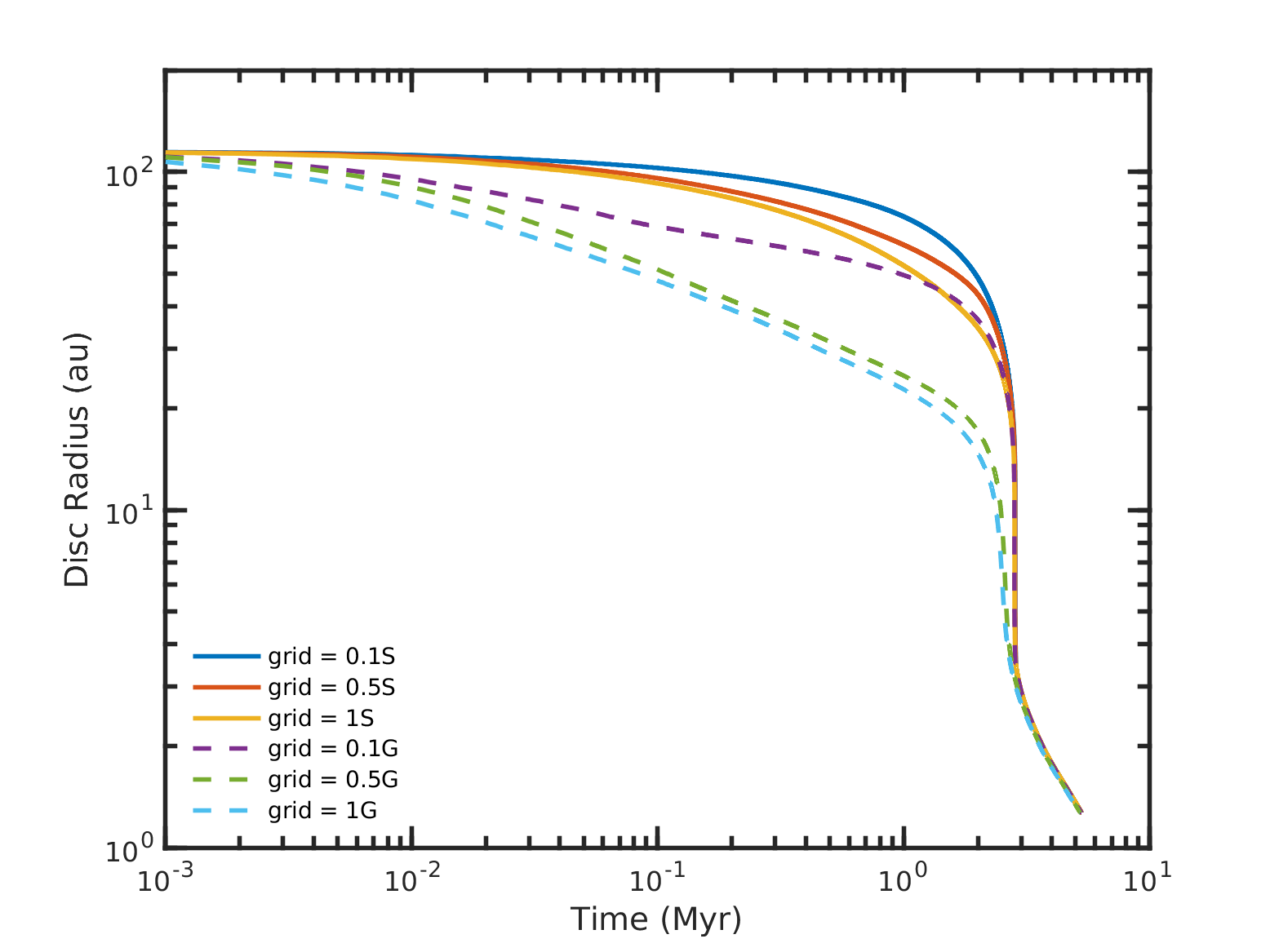}
\caption{Temporal evolution of the disc radius containing 90\% of the mass for discs using the different subgrids of {\sc fried}. The specific subgrid is shown in the legend, with the $\rm f_{PAH}$ ratio of between 0.1, 0.5 and 1. The letters `S' and `G' denote whether the grid is using ISM-like dust, and with the inclusion of grain growth respectively. Solid lines show discs where the dust is ISM-like and dashed lines show those discs where grain growth is assumed to have occurred.}
\label{fig:r90_sub_grid}
\end{figure}

To highlight the effects of the different microphysics in the {\sc fried} subgrids, Fig. \ref{fig:r90_sub_grid} shows the evolution of the disc radii, taken as the radius that encompasses 90\% of the disc mass, with the colours showing the different {\sc fried} subgrid.
The discs simulated here were in an intermediate UV environment of $10^3 \rm G_0$, with $\alpha=10^{-3.5}$ and the central X-ray luminosity $L_X=3\times 10^{30} \rm erg~s^{-1}$.
Solid lines show the subgrids that did not assume any grain growth with the dust being ISM-like, whilst dashed lines show the subgrids that assumed grain growth had occurred.
As can be seen by the solid lines, those discs with ISM-like dust evolved very slowly, even in such a strong UV environment. This is in stark contrast to those discs where grain growth was assumed to have occurred, with the discs reducing in size at a much faster rate. This shows the influence that the reduction in cooling from larger grains has on the mass loss rates, as more gas is able to to be heated and launched in the wind. These differences are observed throughout most of the disc lifetime, until internal photoevaporation opens a gap in the discs and allows the outer disc to be rapidly dispersed. This is shown by the substantial drop in disc radii in all of the evolving discs at $\sim3$Myr. Throughout most of the disc lifetime, the differences in disc radii between those discs evolving with ISM-like dust and those where grain growth has assumed to occur, is $\sim 50\au$, highlighting the importance of understanding the size and role of the dust in external photoevaporative winds.

While grain growth in the disc and the associated entrainment of dust in the wind has been shown to significantly affect the mass loss rates, the other parameter that affects the mass loss rates within the {\sc fried} grids is the PAH-to-dust abundance. In Fig. \ref{fig:r90_sub_grid} the effects of PAH abundance is shown by the different coloured lines for the solid and dashed lines. Whilst their effects on the disc evolution are not as evident as that due to grain growth in the disc/dust depletion in the wind, there are significant effects nonetheless. For example, comparing the blue solid to the yellow solid lines, showing the radius evolution for discs with $f_{\rm PAH}=1$ and $f_{\rm PAH}=0.1$ respectively, there is a difference of up $30\au$ over the course of the disc lifetime. Even though this is not as large as the $\sim 50\au$ seen for the change in dust size, this does add additional complications when exploring the evolution of protoplanetary discs. With very few observational constraints on the PAH abundance available \citep{Berne12,Vicente13}, it is difficult to observationally constrain this quantity, and so further observations would be useful in determining the effects of the PAH abundance on protoplanetary disc evolution.

What the differences in the evolution of the discs whilst using different subgrids of {\sc fried} show, is the importance of understanding the microphysical processes that affect the mass loss rates through external photoevaporation. This includes obtaining constraints from observations of for example PAH abundances, or the presence of grain growth.
It is obvious here that when changing the details of the microphysics, this has a significant effect on the evolution of disc radii, and subsequently the disc mass.
However, Fig. \ref{fig:r90_sub_grid} only shows the evolution for discs in a strong environment, and with single values for the X-ray luminosity of the star and the level of turbulence in the disc.
Next we will examine the effects that these other parameters have on the evolution of protoplanetary discs.

\begin{figure}
\centering
\includegraphics[scale=0.4]{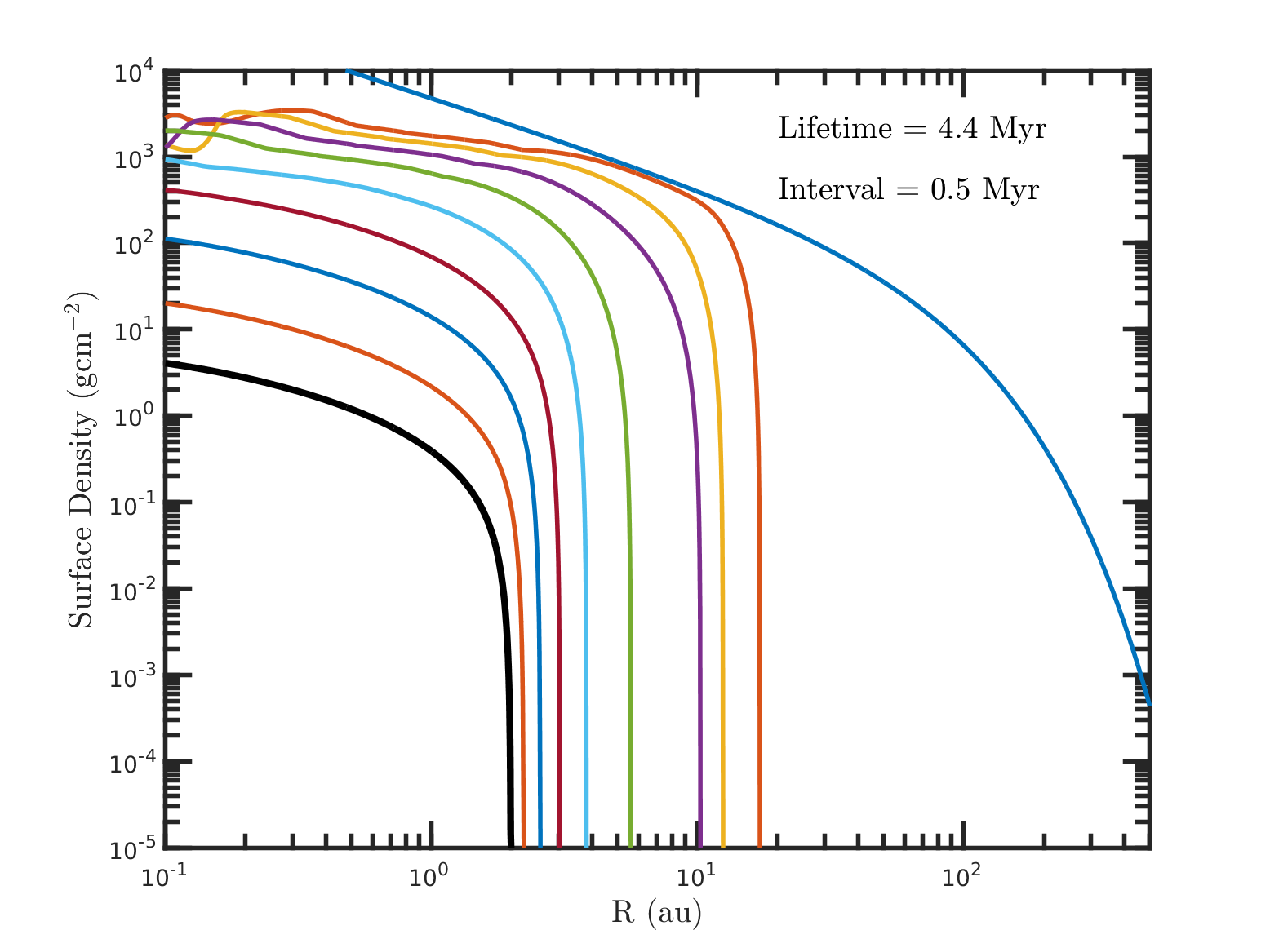}
\caption{Same as Fig. \ref{fig:disc_1000G0_1G} but for a disc in a strong UV environment ($10^5 \rm G_0$).}
\label{fig:disc_10_5G0_1G}
\end{figure}

\subsection{Impact of microphysics as a function of ambient UV field strength}

Previous works have found that the local environment can have significant effects on the evolution of protoplanetary discs, including not only their morphology, but also the time in which they evolve \citep{Coleman20,Coleman22}.
Here we examine the influence that the strength of the external environment (i.e. the star forming region) has on disc evolution with the new {\sc fried} grid.
Whilst Fig. \ref{fig:disc_1000G0_1G} showed the evolution of a protoplanetary disc in an intermediate environment ($10^3 \rm G_0$), Fig. \ref{fig:disc_10_5G0_1G} shows the evolution of a disc in a strong environment, $10^5 \rm G_0$, where the discs are evolving much closer to more massive O-type stars. Here we only show the surface density evolution for the subgrid 1--G.
Whilst the general evolution appears similar, with the discs truncating, and reducing in size and mass over time, there are some notable differences.
In Fig. \ref{fig:disc_1000G0_1G}, the disc was truncated to $\sim 70\au$ after around 2 Myr. With the stronger FUV field, arising from the stronger environment, this was not the case in Fig. \ref{fig:disc_10_5G0_1G}, and in fact the disc had already been truncated to $15 \au$ after only 0.5 Myr, as a result of the larger mass loss rates associated with being situated closer to more massive stars in a stronger FUV environment. Once the disc was effectively truncated, the mass loss rate dropped to being negligible since the energy imparted on to the gas by the UV photons was not strong enough to liberate them from the system. The outer edge of the disc is then determined between the equilibrium between outward viscous spreading, and the weak photoevaporative mass loss rates from both internal and external sources, with both being weak in this region of the disc. The other notable difference is that the lifetime of the disc was significantly shorter in the stronger environment, 4.3 Myr, versus 5.2 Myr. This shorter lifetime being a result of the outer disc being very quickly evaporated away, substantially reducing the supply of gas to the inner disc for it to feed on to the central star.
It is still fairly long by protoplanetary disc standards \citep{Haisch01,Ribas15}, mainly due to the viscous time-scale to remove the inner disc, and so with this being dependent on $\alpha$, a different value there would substantially affect this lifetime.

\begin{figure}
\centering
\includegraphics[scale=0.4]{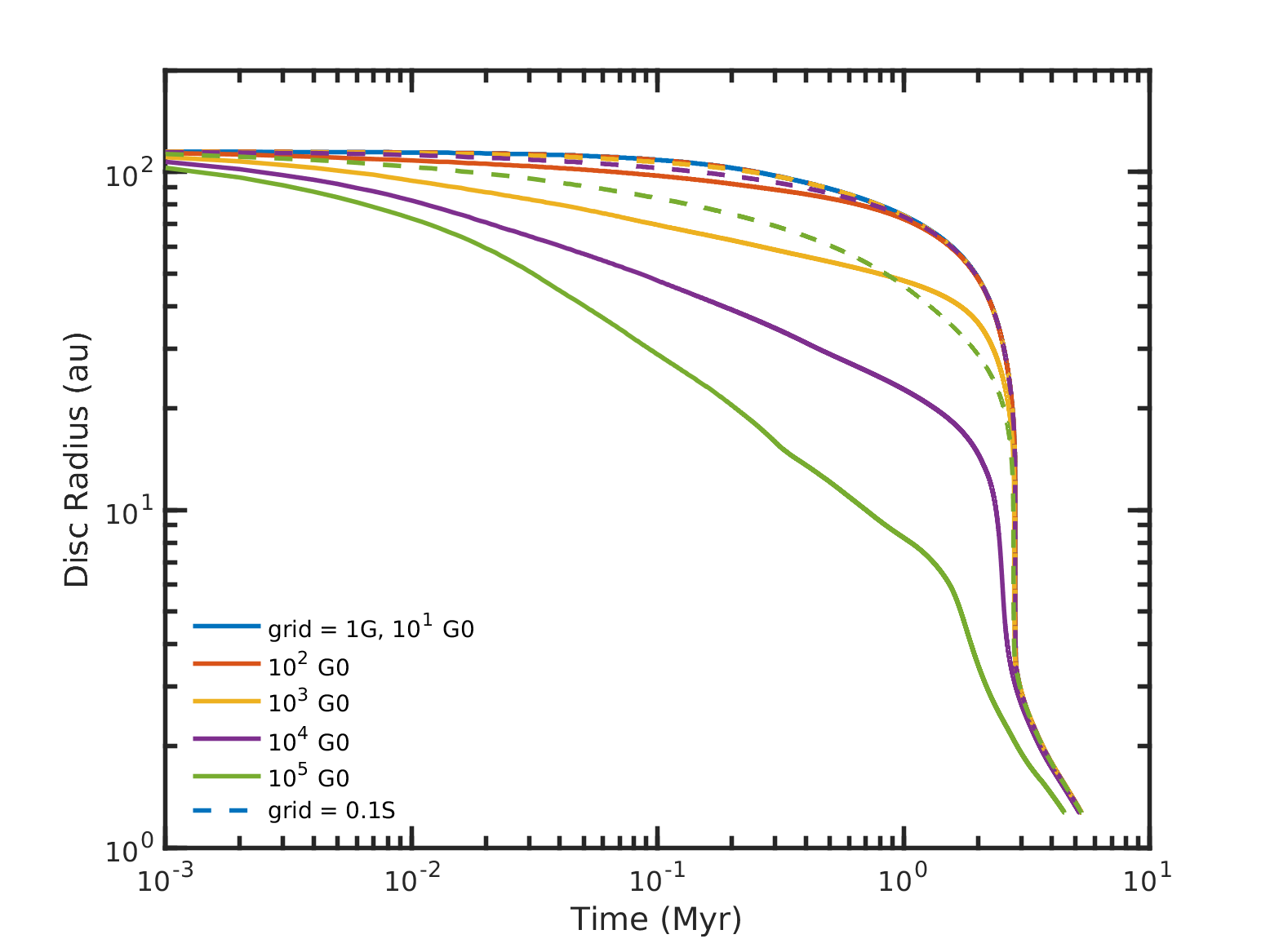}
\caption{Temporal evolution of the disc radius containing 90\% of the mass for discs in different UV environments, ranging from $10^1$ (blue lines) to $10^5$ (green lines). Solid lines show the {\sc fried} subgrid of $\rm f_{PAH}=1$ and including grain growth, whilst dashed lines show the subgrid of $\rm f_{PAH}=0.1$ and only ISM-like dust.}
\label{fig:r90_G0}
\end{figure}

\begin{figure}
\centering
\includegraphics[scale=0.4]{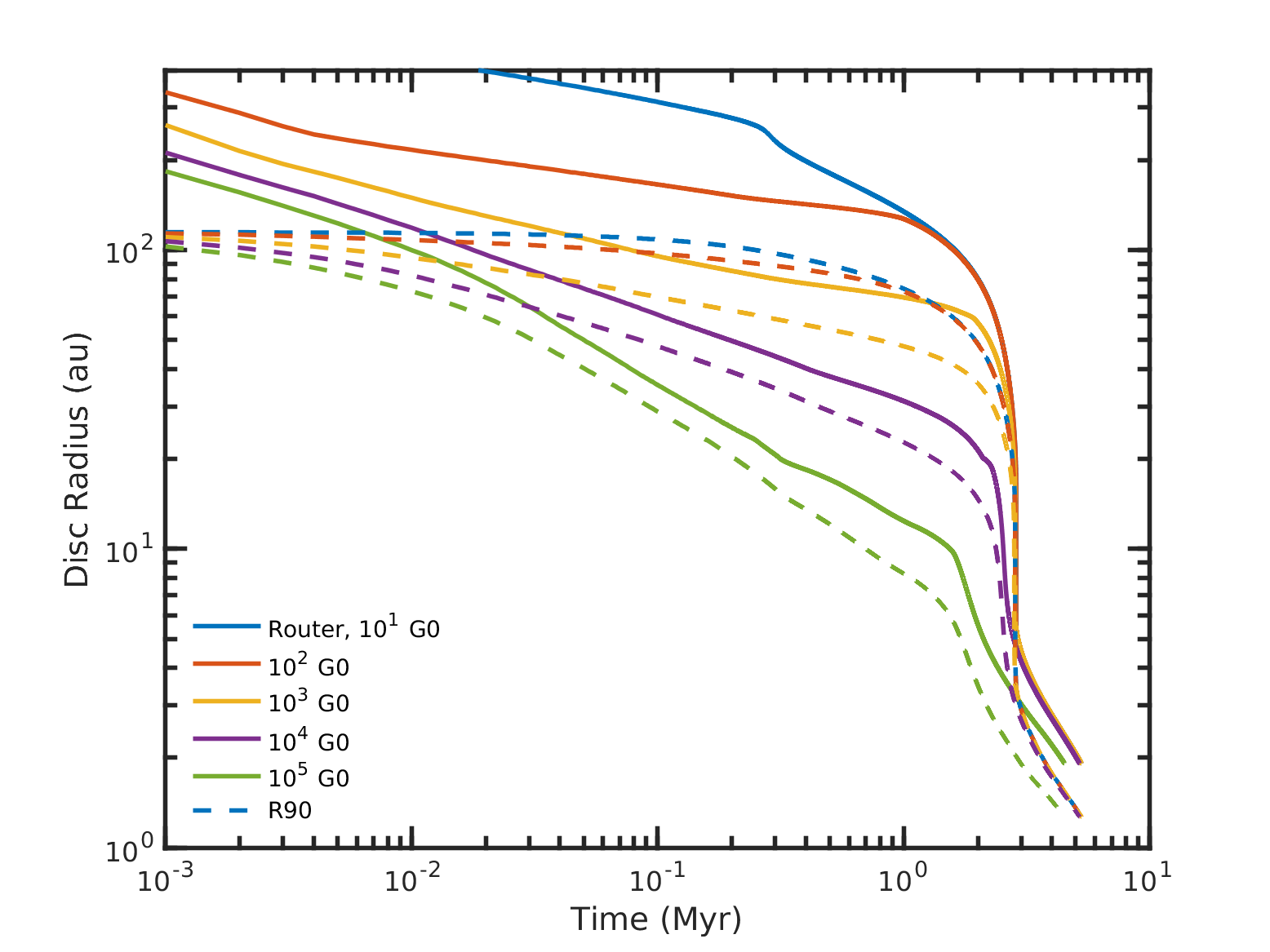}
\caption{Temporal evolution of the disc radius for discs in different UV environments using the {\sc fried} subgrid 1--G, ranging from $10^1$ (blue lines) to $10^5$ (green lines). Dashed lines show the disc radius as calculated for that containing 90\% of the mass, whilst solid lines show the outer edge of the disc where the surface density drops below $10^{-5}\rm gcm^{-2}$.}
\label{fig:r90_router}
\end{figure}

\subsubsection{Effects on disc radii}

Whilst it is interesting to see the evolution of the disc surface density profiles, one measure that can be used to ascertain the effects of the external environment, is the temporal evolution of the disc radius. The disc radius in some given tracer (e.g. continuum, CO) is usually calculated observationally as that containing 90\,per cent of the flux  \citep[e.g.][]{Ansdell18, Trapman2020, Trapman2023}. We therefore use the radius containing 90\,per cent of the disc mass when discussing the disc radius here. This is more likely to be comparable to the gas disc tracing CO radius (as defined above) than continuum given that gas discs are usually more extended \citep[e.g.][]{Ansdell18, Trapman2019}, though in high UV environments the gas-to-dust disc radii are more comparable \citep{Boyden2020}.
Figure \ref{fig:r90_G0} shows the evolution of the disc radius, taken where 90\% of the mass is interior, with the different colours denoting the strength of the external environment, ranging from $10^1$--$10^5$. The solid lines show the temporal evolution for discs using the {\sc fried} subgrid with $\rm f_{\rm PAH}=1$ and including the effects of grain growth(1--G), with the dashed lines being for the subgrid with $\rm f_{\rm PAH}=0.1$ and with the dust being ISM-like(0.1--S).
The speed at which the disc is truncated in the strong environment with $10^5 \rm G_0$ is clearly shown by the solid green line.
After 1 Myr of evolution, the disc can be seen to be truncated to less than 10$\au$.
As the strength of the environment decreases, i.e. going to weaker $\rm G_0$ values, the speed of truncation decreases. However, even in a moderate environment, e.g. $10^3 \rm G_0$, protoplanetary discs are quickly truncated to $\sim70 \au$ and only truncate further over time.

As shown in \cite{Haworth23}, the subgrid with $\rm f_{PAH}=1$ and including the effects of grain growth yielded the strongest mass loss rates. Conversely, the subgrid with $\rm f_{PAH}=0.1$ and only being comprised of small ISM-like dust yielded the weakest mass loss rates.
In Fig. \ref{fig:r90_G0}, the dashed lines show the evolution of the disc radius for the weakest subgrid of {\sc fried}. As can be seen, the discs remain much more extended, and subsequently more massive for longer, as the weaker mass loss rates struggle to disperse the outer disc.
Only when the UV field is extremely strong, is the disc radius significantly affected by the external UV radiation field. For the other cases, they all follow similar evolution tracks.
It is worth noting that even though for all discs with both the strongest and weakest subgrids, weak UV fields do not obviously affect the evolution of the radius that contains 90\% of the mass, they do actually affect the outer regions of the disc.

Figure \ref{fig:r90_router} shows the differences between the radius containing 90\% of the mass (dashed lines), and the outer radius taken where the surface density drops below our numerical floor value\footnote{We note that using surfacer densities of less than $10^{-4}\rm gcm^{-2}$ should also suffice since the surface density at the disc outer edge quickly drops to zero when they are affected by external photoevaporation.} of $10^{-5}\rm gcm^{-2}$ (solid lines), with the colours showing the strength of the external environment.
For the stronger environments, where the UV field is stronger than $10^3 \rm G_0$, there is a minimal difference between the outer radius of the disc, and the radius containing 90$\%$ of the mass.
This is due to the sharpness of the drop in surface density as seen for example by the outer edges of the surface density profiles presented in Fig. \ref{fig:disc_10_5G0_1G}.
For the weaker environments shown by the blue and red lines, there is a much larger difference between the methods of determining the disc radius, since viscous spreading is balancing external photoevaporation at a larger disc radius, causing the disc to have a shallower exponential tail.
But, as the blue solid line shows, the disc is constantly reducing in size, as external photoevaporation is acting to truncate the disc.
Indeed, after 1 Myr the outer disc radius for the disc shown by the blue solid line is at 134 $\au$, showing that even in weak UV environments, external photoevaporation plays a prominent role in the evolution of the outer regions of the discs and cannot be ignored in determining their evolution.

\begin{figure}
\centering
\includegraphics[scale=0.55]{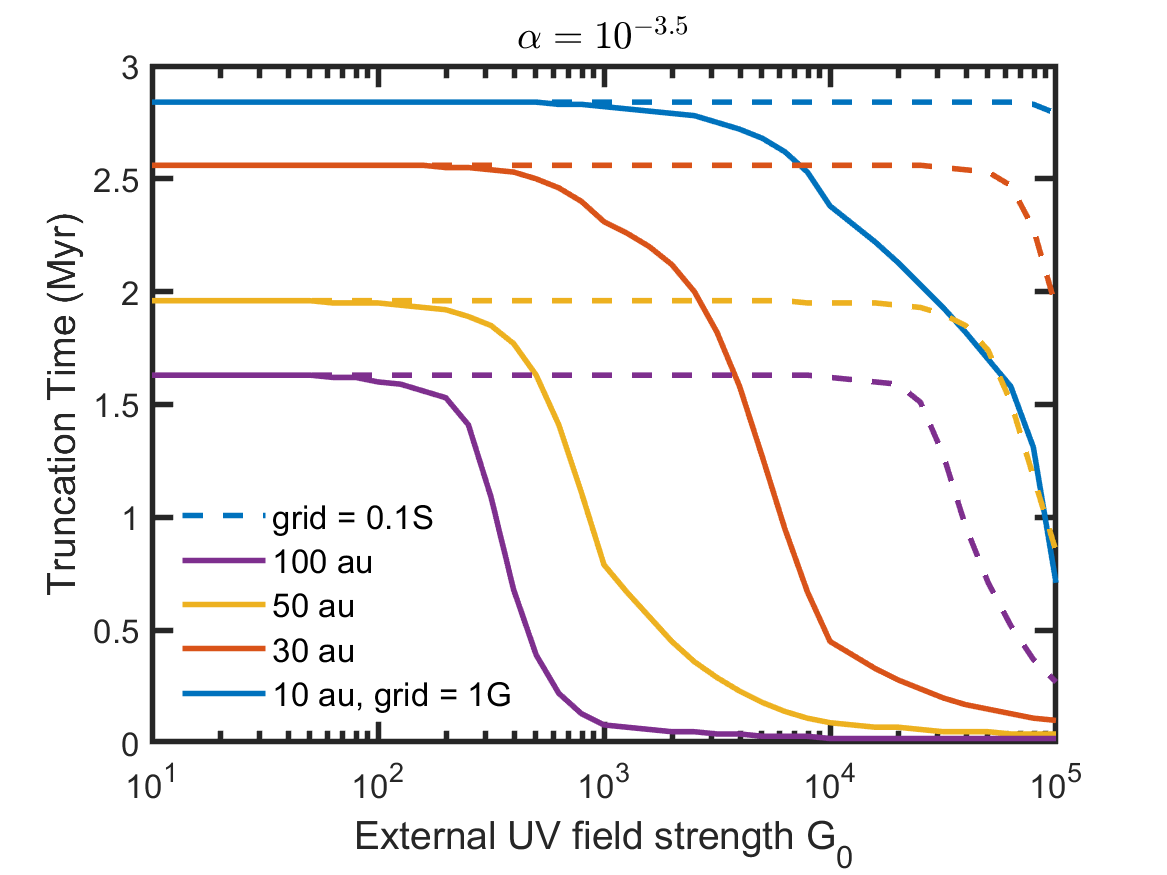}
\caption{Truncation time of protoplanetary discs in different UV environments. Colours show truncation to different disc radii: 10$\au$ (blue), 30$\au$ (red), 50$\au$ (yellow), 100$\au$ (purple). Radii values are taken as the outer radius for the purple line denoting where that falls below 100$\au$, with the radius containing 90\% of the disc mass taken for the other lines, so they are more comparable with observations. Solid lines show truncation times for the subgrid 1--G, whilst dashed lines show for subgrid 0.1--S.}
\label{fig:uv_time}
\end{figure}

With truncation of the outer disc radius occurring for all discs, Fig. \ref{fig:uv_time} shows the truncation times for discs in different UV environments, with the colours showing the times to different disc radii, including 10$\au$ (blue), 30$\au$ (red), 50$\au$ (yellow) and 100$\au$ (purple). We use the outer radius value for the purple line showing 100$\au$ since 90\% of the disc mass is always within 100$\au$ for our discs. For the other radii, we take the radius that contains 90\% of the disc mass to be more comparable with observations.
Solid lines show the results for the subgrid 1--G, with dashed lines showing for 0.1--S.
For the other parameters here, $\alpha=10^{-3.5}$ and $L_{\rm X}=3\times10^{30} \rm erg~s^{-1}$.
The effect of a strong environment is clear when looking at the drops in truncation times for discs of different radii, especially when using the subgrid 1--G.
Looking at the red solid line showing the truncation time down to 30$\au$, strong environments with UV fields $>10^{4}\rm G_0$ drive strong mass loss rates leaving the discs truncated down to that radius after only $\sim 0.5$ Myr. For even stronger environments, truncation is down to 10$\au$ after that time, as shown by the blue solid line.
As can be seen on the left side of Fig. \ref{fig:uv_time}, weak UV environments ($\le 400 \rm G_0$) truncate the discs to specific radii at similar times, for example the yellow solid line shows that truncation down to 50$\au$ occurs after $\sim$2 Myr in these environments.
Whilst this may show that external photoevaporation in weak UV environments has a limited effect on the discs out to 100$\au$, as was shown in Fig. \ref{fig:r90_router} even the weakest environments act to truncate the disc outer edges down to $\sim$few hundreds $\au$, and so their effects can not be neglected in disc evolutionary studies, and after less than 2 Myr, even they have truncated the discs fully down to less than 100$\au$.
Interestingly, the effect of the choice of the {\sc fried} subgrid can also be seen in Fig. \ref{fig:uv_time}, with the weakest subgrid shown by the dashed lines, putting very little environmental dependence on truncation times, apart from in the strongest environments.
Given that the difference in the subgrids is based on the microphysics that determine the mass loss rates, more specifically the PAH-to-gas ratios and whether substantial grain growth has occurred, the stark differences in how the discs evolve highlights that it is important to obtain observations that can provide constraints on the parameters that dominate the microphysical processes for external photoevaporation.

\begin{figure}
\centering
\includegraphics[scale=0.6]{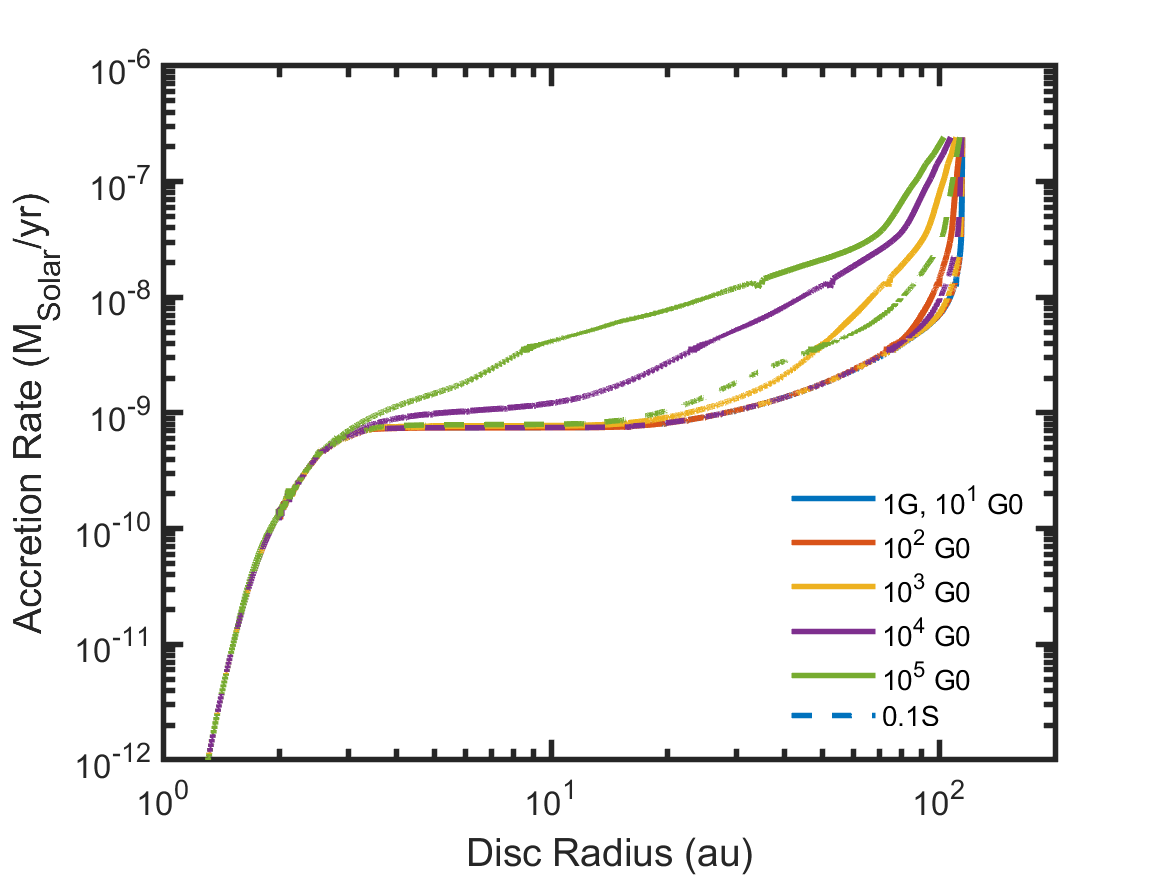}
\caption{Temporal evolution of the disc radius containing 90\% of the mass for discs using the different subgrids of {\sc fried}. The specific subgrid is shown in the legend, with the $\rm f_{PAH}$ ratio of between 0.1, 0.5 and 1. The letters `S' and `G' denote whether the grid is using ISM-like dust, and with the inclusion of grain growth respectively. Solid lines show discs in a strong UV environment ($10^4 \rm G_0$), whilst dashed lines show discs in a weak environment ($10^2 \rm G_0$).}
\label{fig:r90_mdot}
\end{figure}

\subsubsection{Variations in mass accretion rates}

Whilst we have explored the evolution of the radial extent of discs in different environments, another measurable observed quantity is the mass accretion rate on to the central star.
In Fig. \ref{fig:r90_mdot} we show the mass accretion rate on to the central stars as a function of the disc size (taken at 90\% of the disc mass), for discs evolving in different environments. As before the colours show the strength of the external UV field, and solid lines show the discs evolving with the subgrid 1--G, with dashed lines showing for subgrid 0.1--S.
It is clear that for the subgrid 1--G, the strength of the environment has a large impact when comparing these two observed quantities. The mass accretion rate on to the star is determined by the inner disc properties, and so should not be affected by external photoevaporation, until later in the disc lifetime when the the supply to the inner disc is reduced in stronger environments. However with external photoevaporation affecting the disc radius, as seen above, this allows more compact discs to have higher accretion rates, purely because they are younger and the inner disc has not yet been able to accrete on to the star and reduce its accretion rate. This is especially seen in strong environments, shown by the green line, that has mass accretion rates $\sim$ an order of magnitude larger than those in weaker environments, i.e. the yellow line, for a wide range of disc sizes. Such a difference when comparing mass accretion rates to disc sizes, should be observed when exploring different star forming regions, e.g. Lupus or Orion, where the median UV strength significantly differ from $\sim$few $\rm G_0$ in Lupus to $\sim 10^{2}$--$10^{3}\rm G_0$ for $\sigma$ Ori to $>10^{4}\rm G_0$ for ONC \citep{Anania25}.
Since, as shown above, the weaker subgrid of {\sc fried}, 0.1--S, has much weaker dependence on the external environment, this effect is not as obvious, except for in the most strongest of environments. This also means that if observations were able to show differences in this metric for different star forming regions, then some further constraints could be placed on the microphysics of external photoevaporation, purely from an indirect disc evolution viewpoint, rather than by direct measurements of the properties themselves.

\begin{figure*}
\centering
\includegraphics[scale=0.55]{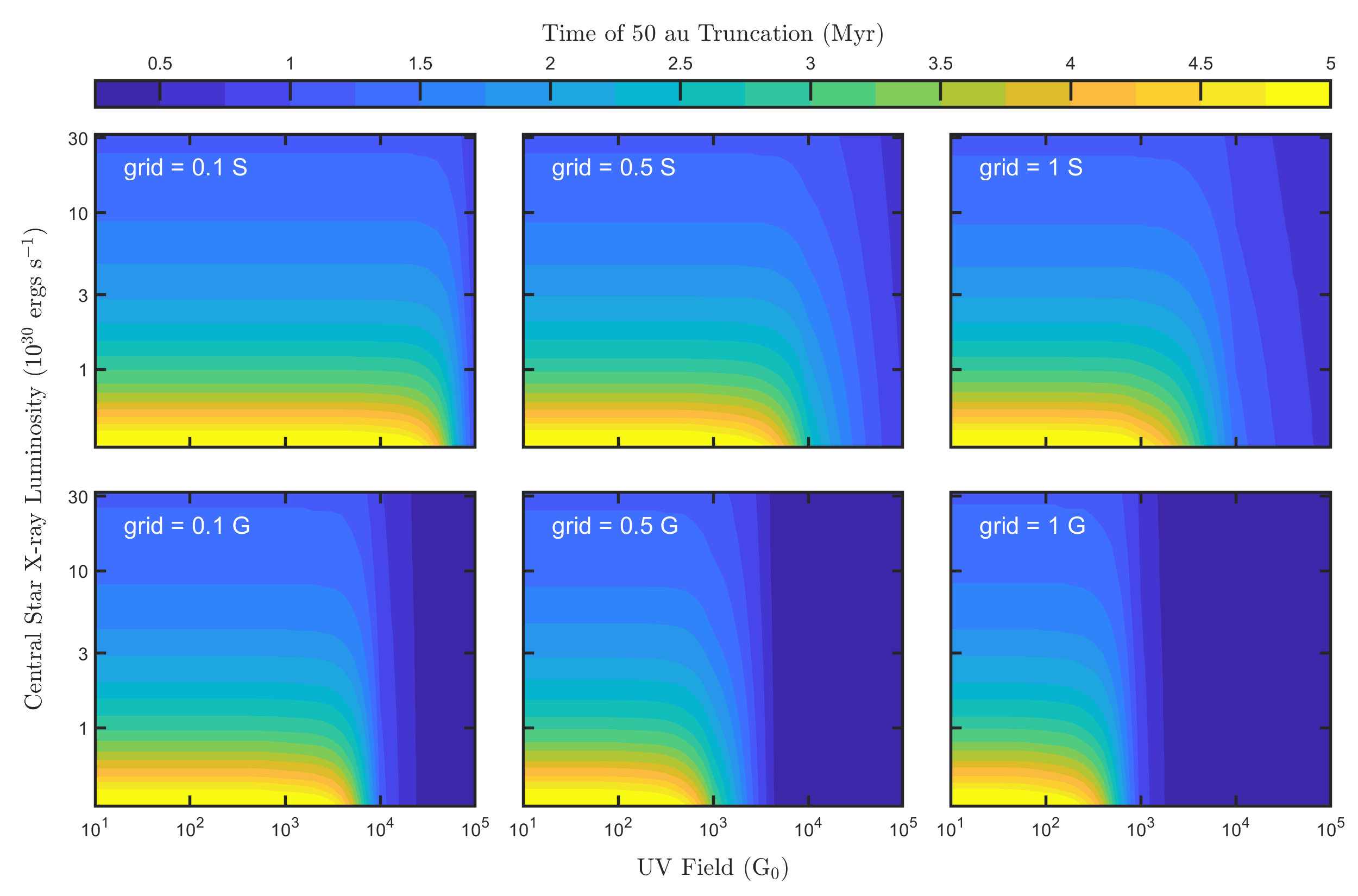}
\caption{Contour plots showing the time taken for discs to be truncated to 50 $\au$ when evolving in different external UV environments (x-axes), different stellar X-ray luminosities (y-axes) and with different subgrids of {\sc fried} (individual panels). Note that the subgrids with ISM-like dust are on the top panels, whilst those that assume grain growth are in the bottom panels. The PAH-to-dust ratios are increasing towards the right.}
\label{fig:50au_trunc}
\end{figure*}

\subsection{Importance for comparing external photoevaporation to other mass loss mechanisms}

Whilst external photoevaporation plays an important role in the evolution of protoplanetary discs, other mechanisms are also responsible for removing mass from the disc. These are mainly internal photoevaporation, and accretion of gas on to the central star through either viscous transport or MHD winds \cite{Coleman24MHD}. Previous works have shown that external and internal photoevaporation are the dominant dispersal mechanisms for protoplanetary discs, with their relative strengths determining which one dominates the evolution of the discs \citep{Coleman22}.

The previous subsections explored the effects of the different \textsc{fried} subgrids on the external photoevaporation rates, and their subsequent effects on the evolution of protoplanetary discs. Those models included only a single value for the strength of the central star's X-ray luminosity, that drives internal photoevaporation. We now expand on this and explore the influence of the subgrids on discs with different strengths of internal photoevaporation to determine where the different mass loss mechanisms dominate each other, when taking the microphysics of external photoevaporation into account.
To do this, in Fig. \ref{fig:50au_trunc} we show contour plots for the truncation time of the disc to 50 $\au$ for discs experiencing different internal and external photoevaporation rates ($y$ and $x$-axis of each panel respectively), and with different {\sc fried} subgrids (different panels).
Since the effects of the viscous parameter $\alpha$ mainly affect the accretion on to the central star, and are typically secondary to photoevaporation in the outer disc regions \citep{Coleman24MHD}, we only show here contour plots for discs with $\alpha=10^{-3.5}$, and note that similar contour plots were found with stronger and weaker $\alpha$ values.
In terms of the {\sc fried} subgrids, the top panels show those discs with ISM-like dust, with the bottom panels showing discs where grain growth is assumed to have taken place. The PAH-to-dust ratio increases from the left to the right of Fig. \ref{fig:50au_trunc}.
Finally, the colour shows the time taken for discs to truncate to 50 $\au$ with blue being extremely quick, and yellow being extremely long.

The effects of a strong UV environment is clear in the bottom panels of Fig. \ref{fig:50au_trunc} where irrespective of the PAH-to-dust ratio and the strength of the central star's X-ray luminosity, discs are quickly truncated to $50 \au$ when the UV environment is at least $10^4 \rm G_0$. Only for the most effective subgrid, 1--G, is the truncation efficient down to $10^3 \rm G_0$. For weaker environments, there appears to be very few differences when changing the UV field strength. However for these discs, the effects of internal photoevaporation can be seen as the truncation time decreases as $L_X$ increases. This is due to internal photoevaporation removing the intermediate regions of the disc, reducing the resupply rate of the outer disc that allows it to reduce in mass and truncate at a faster rate.

When comparing the top to the bottom panels of Fig. \ref{fig:50au_trunc} the effect of choosing which grid, especially on how much grain growth has occurred becomes noticeably apparent. Whilst the effects of internal photoevaporation are still visible, those of external photoevaporation are less so with only the strongest environments now being able to truncate the discs to 50$\au$ within 1 Myr. Looking at the top left panel, showing the weakest rates that arise from the {\sc fried} subgrid, there is very little dependence on external photoevaporation. This again highlights the importance of placing constraints on parameters that affect the microphysics of external photoevaporation, namely the efficiency of grain growth and the PAH-to-dust ratios.

\subsection{Outlook for determining microphysics parameters}
In this paper we have demonstrated that the microphysical properties related to external photoevaporation can affect the evolution of macroscopic disc properties. This further motivates efforts to determine the dust and PAH properties in the winds of externally photoevaporating discs. Fortunately, with JWST we are now in an era where determining these parameters for a statistically significant sample is now possible. JWST is sensitive to spectroscopic PAH features such as the 3.3 and 3.4\,$\mu$m C-H vibrational bands of PAH's with NIRSpec \citep[e.g.][]{2024Sci...383..988B} and multiple features, including at 7.7, 8.4 and 11.27\,$\mu$m with MIRI \citep[e.g.][]{2024A&A...689A..92S}. JWST can also provide information on the dust and ice properties \citep[even with NIRCam, e.g.][]{2025ApJ...979..110B}. For evaporating discs in silhouette against photoionised gas, the dust properties in the wind can also be probed using the attenuation of nebular emission relative to the surrounding H\,\textsc{ii} region \citep[e.g.][]{2012ApJ...757...78M}. Spectroscopic JWST data already exists for a number of proplyds in the Trapezium Cluster, as well as imaging in multiple bands across the region \citep{2022PASP..134e4301B, 2024A&A...685A..73H, 2024A&A...685A..74P, 2024A&A...685A..77P, 2023arXiv231003552M} and so PAH and dust abundances in winds should follow soon.

\section{Summary and Conclusions}
\label{sec:conclusions}
Recent work has presented updated mass loss rates that arise from external photoevaporation \citep[{\sc fried},][]{Haworth23}. In this work we explore the different variations in the {\sc fried} grid that arise due to different treatments of the microphysics within external photoevaporation, namely the assumption on dust growth and the PAH-to-dust ratios. Additionally we also explore the effects that the external environment, central star's X-ray luminosity, and the strength of turbulence, have on the evolution of protoplanetary discs. We draw the following main conclusions from this work.\\

1. The treatment of the microphysics that affects external photoevaporation has a profound effect on the strength of external mass-loss rates. These effects mainly arise due to whether grain growth has occurred in the disc, reducing the amount of radiation absorbed by dust and increasing the rates at which gas is lost from the discs. Additionally, having larger (ISM-like) PAH-to-dust ratios further increases the mass loss rate, but to a lesser extent than the assumption of dust grain sizes. When both of these microphysical assumptions are made, and there is a sufficient strength of external radiation (>100 $\rm G_0$) the rate of evolution of protoplanetary discs, especially their radii, are substantially enhanced.\\

2. The external environment is also found to have a profound effect on the evolution of protoplanetary discs, consistent with previous studies \citep{Coleman22}. This is mainly seen in the stronger subgrids of {\sc fried} that included grain growth and had larger (ISM-like) PAH-to-dust ratios. Indeed when using the subgrid 1--G, discs in strong UV environments were rapidly truncated on 0.1--1 Myr time-scales. In contrast, the subgrid with ISM-like dust and reduced PAH-to-dust ratios showed very little dependence on the the external environment. 
It is also interesting to note that even in weak environments, even though discs did not truncate efficiently, their outer edges were affected and maintained at a truncated radius, typically $\sim$few hundred $\au$.\\

3. When combining mass accretion rates on to the central star with the disc radius, we find that discs evolving in different strength environments exhibit different relationships between the accretion rate and measured disc radius. The accretion rate is mainly determined by the evolution of the inner disc of the system, which for a significant amount of the disc lifetime can be detached from the outer disc, i.e. it is only affected when the supply of the outer disc to the inner disc is sufficiently reduced. This leads to a signature of external photoevaporation when comparing discs of similar sizes in different regions, since those evolving in the stronger environment would have larger mass accretion rates than their weaker environment counterparts. Effectively, this also also traces a rough relative age of the discs in these regions, since the outer disc radius is inexplicably linked to the truncation time for discs in specific UV field strengths.\\

4. Whilst we find that the microphysics of external photoevaporation, and the strengths of both internal and external photoevaporation, have a large range of effects on the evolution of protoplanetary discs, the strength of the turbulence parameter $\alpha$ is found to have very little effect. This is due to the angular momentum transfer time-scales in the outer disc being very long and so the evolution of those regions is mainly dictated by photoevaporation. Only through the resupply of material that is lost to wind, is there any small influence from $\alpha$. For the inner disc on the other hand, since photoevaporation is ineffective here, the inner discs lifetimes are comparable to the viscous time-scale and so are directly affected by $\alpha$. This leads to the main role that $\alpha$ has in determining the disc evolution being mainly the final lifetime of the inner disc and accretion on to the central star, which are largely unaffected by the other parameters studied here that mainly affect the outer disc regions. \\

Overall, this work shows the evolution of protoplanetary discs depends significantly on the strength of photoevaporative mass loss rates. More importantly the specific details of the photoevaporation processes are important as they can drastically change the effectiveness of external photoevaporative mass loss rates. Whether dust grains are treated as ISM-like or if grain growth has occurred is seen to have the largest effect on disc evolution. Additionally the PAH abundance is also found to affect disc evolution, with more ISM-like PAH abundances leading to faster evolution. What these results show is that all of these processes are important in disc evolutionary studies and should be included when comparing theoretical models with observations. Furthermore, they highlight the need to obtain more stringent constraints on the underlying physics that affect disc evolution, including the sizes of grain in photoevaporative winds, and the PAH abundance in discs. JWST should soon provide these constraints. 

With photoevaporation being the main driver of the mass that is lost in protoplanetary discs, it makes it extremely difficult to distinguish between numerous methods for angular momentum transfer that drives accretion on to the central stars. Whilst our results don't heavily depend on the level of viscosity did not have a significant effect on the disc properties, apart from their final lifetimes, we did not include models of MHD driven winds that are put forward as a driver of angular momentum transfer \citep{Tabone22}. However, recent work has shown that when including internal and external photoevaporative winds within viscous and/or MHD wind driven models, most signatures of differences between the drivers of angular momentum transfer are washed out \citep{Coleman24MHD}. It remains to be seen whether if observed discs obtained precise enough constraints on the local environment and stellar properties, as well as the properties of the microphysics discussed above, then would it be possible to distinguish between viscous and MHD wind driven processes. Additionally, with large populations of discs, such differences may arise from a statistical standpoint, as was shown by \citet{Alexander23} between wind driven discs and viscously evolving discs including internal photoevaporation. This will be explored in future work, since as most planets are thought to form within protoplanetary discs, the lifetime and evolution of those discs will significantly affect the formation and evolution of planets and planetary systems.

\section*{Data Availability}
The data underlying this article will be shared on reasonable request to the corresponding author.

\section*{Acknowledgements}
The authors thank the anonymous referee for providing useful and interesting comments that improved the paper.
GALC acknowledges funding from the Royal Society under the Dorothy Hodgkin Fellowship of T. J. Haworth, and UKRI/STFC grant ST/X000931/1.
TJH acknowledges funding from a Royal Society Dorothy Hodgkin Fellowship, which at the time of submission also funds GALC. TJH also acknowledges UKRI guaranteed funding for a Horizon Europe ERC consolidator grant (EP/Y024710/1).
This research utilised Queen Mary's Apocrita HPC facility, supported by QMUL Research-IT (http://doi.org/10.5281/zenodo.438045).

\vspace{-0.2cm}
\bibliographystyle{mnras}
\bibliography{references}{}

\vspace{-0.2cm}
\label{lastpage}
\end{document}